\documentclass{article}
\usepackage{amscd,amsmath,amssymb,bbm}
\usepackage{amsfonts}
\newcommand{\p}[1]{(\ref{#1})}

\newcommand{\bD}{\overline{D}{}}
\newcommand{\bLam}{{\overline{\Lambda}}{}}

\newcommand{\bxi}{{\bar\xi}}
\newcommand{\blam}{{\bar\lambda}}

\newcommand{\brho}{{\bar\rho}}
\newcommand{\tb}{{\bar\theta}}
\newcommand{\halpha}{{\hat\alpha}}

\newcommand{\be}{\begin{equation}}
\newcommand{\ee}{\end{equation}}
\newcommand{\bea}{\begin{eqnarray}}
\newcommand{\eea}{\end{eqnarray}}
\newcommand{\ba}{\begin{array}}
\newcommand{\ea}{\end{array}}

\newcommand{\nn}{\nonumber}

\def\im{{\rm i}}

\def\={\ =\ }

\def\Nf{$\cal N${=\,}4~}

\begin{document}
\thispagestyle{empty}
\vspace{2cm}
\begin{flushright}
%Draft ,\; \today
\end{flushright}\vspace{4cm}
\begin{center}
{\Large\bf N=4, d=1 Supersymmetric Hyper-K\"{a}hler Sigma Models \vspace{0.5cm} \\and Non-Abelian Monopole Background}
\end{center}
\vspace{1cm}

\begin{center}
{\Large
Stefano Bellucci$\,{}^{a}$, Sergey Krivonos$\,{}^{b}$  and
Anton Sutulin$\,{}^{b}$
}\\
\vspace{1.0cm}
${}^a$ {\it
INFN-Laboratori Nazionali di Frascati,
Via E. Fermi 40, 00044 Frascati, Italy} \vspace{0.2cm}

${}^b$
{\it Bogoliubov  Laboratory of Theoretical Physics,
JINR, 141980 Dubna, Russia}
\vspace{0.2cm}
\end{center}
\vspace{3cm}

\begin{abstract}
\noindent We construct a Lagrangian formulation of \Nf
supersymmetric mechanics with hyper-K\"{a}hler sigma models
in a bosonic sector in the non-Abelian background gauge field.
The resulting action includes a wide class of \Nf supersymmetric
mechanics describing the motion of an isospin-carrying particle
over spaces with non-trivial geometry. In two examples we discuss in
details, the background fields are identified with the field of
BPST instantons in the flat and Taub-NUT spaces.
\end{abstract}

\newpage
\setcounter{page}{1}

\setcounter{equation}0
\section{Introduction}
The \Nf supersymmetric mechanics provide a nice framework for the
study of many  interesting features of higher dimensional
theories. At the same time, the existence of a variety of
off-shell \Nf irreducible linear supermultiplets in $d=1$
\cite{GR, Tolik, FG, FG1, ikl1} makes the situation in one
dimension even more interesting, and just this prompted us to
investigate such supersymmetric models themselves, without
reference to higher dimensional counterparts. Being a
supersymmetric invariant theory, the \Nf mechanics admits a
natural formulation in terms of superfields living in a standard
and/or in a harmonic superspace \cite{hss}, adapted to one
dimension \cite{IL}. In any case the preferable approach to
discuss supersymmetric mechanics is the Lagrangian one. Being
quite useful, the Lagrangian approach has one subtle point, when
we try to describe the system in an arbitrary gauge background.
While the inclusion of the Abelian gauge background can be done
straightforwardly \cite{IL}, the non-Abelian background asked for
new ingredients - the isospin variables which have to be included
in the description
\cite{FIL,brazil,FIL1,bk,KL1,Kon1,sutulin1,Kon2,Ikon,sutulin2}.
These isospin variables become purely
 internal degrees of freedom after quantization and form an auxiliary
\Nf supermultiplet, together with the auxiliary fermions.

There are different approaches to introduce such  auxiliary
superfields and couplings with them, but up to now all constructed
models have been restricted to have conformally flat sigma models
in the bosonic sector. This restriction has an evident source - it
has been known for a long time that all linear \Nf supermultiplets
can be obtained through a dualization procedure from the \Nf
``root'' supermultiplet  -- \Nf hypermultiplet
\cite{hyper1,hyper2,hyper3,hyper4,root,DI1}, while the bosonic
part of the general hypermultiplet action is conformal to the flat
one. The only way to escape this flatness situation is to use
nonlinear supermultiplets \cite{AS,nelin1,NLM2}, instead of linear
ones.

The main aim of the present paper is to construct the Lagrangian formulation of \Nf supersymmetric mechanics on
conformal to hyper-K\"{a}hler spaces in the non-Abelian background gauge fields. To achieve this goal we combine
two ideas
\begin{itemize}
\item We introduce the coupling of the matter supermultiplets with an auxiliary fermionic supermultiplet $\Psi^{\halpha}$
containing on-shell four physical fermions and four auxiliary bosons playing the role
    of isospin variables. The very specific
    coupling results in a component action which contains only time derivatives of the fermionic components
    present in $\Psi^{\halpha}$. Then, we dualize these fermions into auxiliary ones, ending up with the
    proper action for matter fields and isospin variables. This procedure was developed in \cite{bk}.
\item As the next step, starting from the action for the \Nf tensor supermultiplet \cite{v1,v1a} coupled
with the superfield $\Psi^{\halpha}$, following \cite{AS}, we
dualize the auxiliary component $A$ into a fourth physical boson,
finishing with the action having a geometry conformal to the
hyper-K\"{a}hler one in the bosonic sector.
\end{itemize}
The resulting action contains a wide class of \Nf supersymmetric
mechanics describing the motion of an isospin-carrying particle
over spaces with non-trivial geometry and in the presence of a
non-Abelian background gauge field. In two examples we discussed
in details, these background correspond to the field of the BPST
instanton in the flat and Taub-NUT spaces. In order to make our
presentation self-sufficient, we include in Section 2 a sketchy
description of our construction applied to the linear tensor and
hypermultiplet. We also discuss the relation between these
supermultiplets in the context of our approach (Section 3), which
immediately leads to the generalized procedure presented in
Section 4. \setcounter{equation}0
\section{Isospin particles in conformally flat spaces}
One of the possible ways to incorporate the isospin-like variables
in the Lagrangian of supersymmetric mechanics is to couple the
basic superfields with auxiliary fermionic superfields
$\Psi^\halpha, {\bar \Psi}_\halpha$, which contain these isospin
variables \cite{bk}. Such a coupling, being written in a standard
\Nf superspace, has to be rather special, in order to provide a
kinetic term of the first order in time derivatives for the
isospin variables and to describe the auxiliary fermionic
components present in $\Psi^\halpha, {\bar \Psi}_\halpha$.
Following \cite{bk}, we introduce the coupling of auxiliary $\Psi$
superfields with some arbitrary, for the time being, \Nf
supermultiplet $X$  as \be\label{actionX} S_c=-\frac{1}{32}\int dt
d^4\theta \; \left( X +g\right)  \Psi^\halpha {\bar\Psi}_\halpha,
\qquad g=const. \ee The $\Psi$ supermultiplet is subjected to the
irreducible conditions \cite{ikl1} \be\label{Psi} D^i
\Psi^1=0,\quad D^i \Psi^2+\bD{}^i \Psi^1=0, \quad \bD_i \Psi^2=0,
\ee and thus it contains four fermionic and four bosonic
components \be\label{psi} \psi^\halpha=\Psi^\halpha|, \qquad u^i
=-D^i {\bar\Psi}{}^2|,\quad {\bar u}_i= \bD_i \Psi^1|, \ee where
the symbol $|$ denotes the $\theta=\tb=0$ limit and \Nf covariant
derivatives obey standard relations \be \left\{ D^i,
\bD_j\right\}=2 \im \delta^i_j
\partial_t. \ee It has been demonstrated in \cite{bk} that if the \Nf
superfield $X$ is subjected to the constraints \cite{leva,ikl1}
\be\label{X} D^i D_i X=0,\;\bD_i \bD{}^i X=0,\quad \left[
D^i,\bD_i\right] X=0, \ee then the component action which follows
from \p{actionX} can be written as \bea\label{actionXc} S_c&=
&\int dt\left[ -(x+g)\left({ \rho}{}^1{\brho}{}^2-{
\rho}{}^2{\brho}{}^1\right)-\frac{i}{4} (x+g) \left( {\dot u}{}^i
{\bar u}_i-
u^i\dot{\bar u}_i\right)+\frac{1}{4}A_{ij}u^i{\bar u}{}^j+\right.\nn\\
&& \left. \frac{1}{2}\eta_i\left({\bar u}{}^i \brho{}^2+u^i
\rho{}^2\right)+\frac{1}{2}\bar\eta{}^i\left(u_i \rho{}^1+{\bar
u}_i\brho{}^1\right) \right], \eea where the new fermionic
components $\rho^\halpha,{\bar\rho}_\halpha$  are defined as
\be\label{rho} \rho^\halpha ={\dot\psi}{}^\halpha, \quad
{\bar\rho}_\halpha={\dot\psi}_\halpha. \ee The components of the
superfield $X$ entering the action \p{actionXc} have been
introduced as \be\label{compX}
 x= X|,\; A_{ij} = A_{(ij)}= \frac{1}{2}\left[ D_i,\bD_j\right] X|,\qquad \eta^i= -iD^i X|,\;
\bar\eta{}_i= -i\bD_i X|. \ee What makes the action \p{actionXc}
interesting is that, despite the non-local definition of the
spinors $\rho^\alpha, {\bar\rho}_\alpha$ \p{rho}, the action is
invariant under the following  \Nf supersymmetry transformations:
\bea\label{n4tr1} &&\delta \rho^1=-\bar\epsilon{}^i \dot{\bar
u}_i,\; \delta\rho^2=\epsilon_i\dot{\bar u}{}^i,\quad \delta
u^i=-2i\epsilon^i\brho{}^1+2i\bar\epsilon{}^i\brho{}^2,\; \delta
{\bar u}_i=-2i\epsilon_i\rho{}^1+2i\bar\epsilon_i\rho{}^2, \nn \\
&& \delta x=-i\epsilon_i\eta^i-i\bar\epsilon{}^i\bar\eta_i,\quad
\delta\eta{}^i=-\bar\epsilon{}^i{\dot x}-i\bar\epsilon{}^j
A^i_j,\; \delta \bar\eta_i=-\epsilon_i{\dot x}+i\epsilon_j
A_i^j,\quad \delta A_{ij} = -\epsilon_{(i}\dot\eta_{j)} +
\bar\epsilon_{(i}\dot{\bar\eta}{}_{j)}. \eea In the action
\p{actionXc} the fermionic fields $\rho^\halpha,\brho{}_\halpha$
are auxiliary ones, and thus they can be eliminated  by their
equations of motion \be\label{ad5a}
\rho^1=\frac{1}{2(x+g)}\eta_i{\bar u}{}^i, \qquad
\rho^2=-\frac{1}{2(x+g)}\bar\eta{}^i{\bar u}_i. \ee {}Finally, the
action describing the interaction of $\Psi$ and $X$
supermultiplets acquires a very simple form \be\label{actionXcfin}
S_c=\frac{1}{4}\int dt\left[
 -i (x+g) \left( {\dot u}^i {\bar u}{}_i- u^i\dot{\bar u}{}_i\right)+
 A_{ij}u^i{\bar u}{}^j+
\frac{1}{x+g} \eta_i\bar\eta_j\left( u^i {\bar u}{}^j+ u^j{\bar
u}{}^i\right) \right]. \ee Thus, in the fermionic superfields
$\Psi$ only the bosonic components $u^i,{\bar u}_i$, entering the
action with a kinetic term linear in time-derivatives, survive.
After quantization these variables become purely internal degrees
of freedom.

In order to be meaningful, the action \p{actionX} has to be
extended by the action for the supermultiplet $X$ itself. If the
superfield $X$ obeying \p{X} is considered as an independent
superfield, then the most general action reads \be\label{Xpsi}
S=S_x+S_c=-\frac{1}{32}\int dt d^4\theta {\cal F}(X) +S_c, \ee
where ${\cal F}(X)$ is an arbitrary function of $X$. In this case
the components $A_{ij}$ \p{compX} are auxiliary ones, and they
have to be eliminated by their equations of motion. The resulting
 action describes \Nf supersymmetric mechanics with one physical boson $x$ and four physical fermions
$\eta^i, {\bar\eta}_j$ interacting with isospin variables $u^i, {\bar u}_i$. Just this system has been considered in \cite{{FIL},{FIL1},{bk}}.

It is clear that treating the scalar bosonic superfield $X$ as an
independent one is too restrictive, because the constraints \p{X}
leave in this supermultiplet only one physical bosonic component
$x$, which is not enough to describe the isospin particle. In the
present approach, the way to overcome this limitation was proposed
in \cite{{sutulin1},{sutulin2}}. The key point is to treat the
superfield $X$ as a composite one, constructed from \Nf
supermultiplets with a larger number of physical bosons. The two
reasonable superfields from which it is possible to construct the
superfield $X$ are \Nf tensor supermultiplet ${\cal V}^{ij}$
\cite{{v1},{v1a}} and a one-dimensional hypermultiplet
$Q^{i\alpha}$ \cite{hyper1,hyper2,hyper3,hyper4,IL}.\vspace{0.5cm}

\noindent{\bf Tensor supermultiplet}\\
The \Nf tensor supermultiplet is described by the triplet of
bosonic \Nf superfields ${\cal V}^{ij}={\cal V}^{ij}$ subjected to
the constraints \be\label{V} D^{(i}{\cal V}^{jk)}=\bD{}^{(i}{\cal
V}^{jk)}=0, \qquad \left( {\cal V}^{ij}\right)^\dagger = {\cal
V}_{ij}, \ee which leave in ${\cal V}^{ij}$ the following
independent components: \be\label{v} v^a=-\frac{\im}{2}\left(
\sigma^a\right)_i{}^j{\cal V}_j^i|,\quad
\lambda^i=\frac{1}{3}D^j{\cal V}^i_j|,\quad \blam{}_i
=\frac{1}{3}\bD_j{\cal V}^j_i|, \quad A=\frac{\im}{6}D^i  \bD{}^j
{\cal V}_{ij}|. \ee Thus, its off-shell component field content is
$(3,4,1)$, i.e. three physical $v^a$ and one auxiliary $A$ bosons
and four fermions $\lambda^i, \blam{}_i$ \cite{{v1},{v1a}}. Under
\Nf supersymmetry these components transform as follows:
\bea\label{Vtr} &&\delta v^a=\im \epsilon^i
(\sigma^a)_i^j\blam_j-\im \lambda^i(\sigma^a)_i^j
\bar\epsilon_j,\quad
\delta A=\bar\epsilon_i {\dot\lambda}{}^i-\epsilon^i\dot\blam_i, \nn \\
&&\delta\lambda^i=\im \epsilon^i A+\epsilon^j(\sigma^a)_j^i{\dot
v}{}^a,\quad \delta\blam_i=-\im \bar\epsilon_i
A+(\sigma^a)_i^j\bar\epsilon_j{\dot v}{}^a . \eea Now one may
check that the composite superfield \be\label{XV} X=\frac{1}{|\cal
V|} \equiv \frac{1}{\sqrt{{\cal V}^a {\cal V}^a}}, \ee where
${\cal V}^a=-\frac{\im}{2}\left( \sigma^a\right)_i{}^j{\cal
V}_j^i$, obeys \p{X} in virtue of \p{V}. Clearly, now all
components of the $X$ superfield, i.e. the physical boson $x$,
fermions $\eta^i, \bar\eta_i$ and auxiliary fields $A^{ij}$
\p{compX} are expressed through the components of the ${\cal
V}^{ij}$ supermultiplet \p{v} as \bea\label{rel1} &&
x=\frac{1}{|v|}, \qquad \eta^i = \frac{v^a}{|v|^3}
(\lambda\sigma^a)^i, \quad
\bar\eta_i =   \frac{v^a}{|v|^3} (\sigma^a\blam)_i, \nn \\
&& A^i_j =-3\frac{v^a
v^b}{|v|^5}(\lambda\sigma^a)^i(\sigma^b\blam)_j-
\frac{v^a(\sigma^a)^i_j}{|v|^3}A+\frac{1}{|v|^3}\epsilon^{abc}v^a{\dot
v}{}^b(\sigma^c)^i_j+ \frac{1}{|v|^3}\left( \delta^i_j
\lambda^k\blam_k-\lambda_j\blam^i\right). \eea In what follows we
will also need the expression for $A^i_j$ components \p{rel1} in
terms of  $\eta^i, \bar\eta_i$ fermions \p{compX}, which reads
\be\label{AV} A^i_j =-
\frac{v^a(\sigma^a)^i_j}{|v|^3}A+\frac{1}{|v|^3}\epsilon^{abc}v^a{\dot
v}{}^b(\sigma^c)^i_j-
|v|\left(\eta^i\bar\eta_j+\eta_j\bar\eta{}^i\right)-\frac{1}{|v|}
v^a(\eta\sigma^a\bar\eta)\; v^b (\sigma^b)^i_j. \ee Finally, one
should note that, while dealing with the tensor supermultiplet
${\cal V}^{ij}$, one may generalize the $S_x$ action \p{Xpsi} to
have the full action in the form \be\label{Vpsi}
S=S_v+S_c=-\frac{1}{32}\int dt d^4\theta {\cal F}({\cal V}) +S_c,
\ee where ${\cal F}({\cal V})$ is now an arbitrary function of
${\cal V}{}^{ij}$. After eliminating the auxiliary component $A$
in the component form of  \p{Vpsi} we will obtain the action
describing the \Nf supersymmetric three-dimensional isospin
particle moving in the magnetic field of a Wu-Yang monopole and in
some specific scalar potential \cite{sutulin1}\footnote{An
alternative description of the same system has been recently
constructed in \cite{Ikon}}.\vspace{0.5cm}

\noindent{\bf Hypermultiplet}\\
Similarly to the tensor supermultiplet one may construct the
superfield $X$ starting from the \Nf hypermultiplet. The \Nf, d=1
hypermultiplet is described in \Nf superspace by the quartet of
real \Nf superfields ${\cal Q}^{i\alpha}$ subjected to the
constraints \be\label{Q} D^{(i}{\cal Q}^{j)\alpha}=0, \quad
\bD{}^{(i}{\cal Q}^{j)\alpha}=0,\qquad \left( {\cal
Q}^{i\alpha}\right)^\dagger = {\cal Q}_{i\alpha}. \ee This
supermultiplet describes four physical bosonic and four physical
fermionic  variables \be\label{compQ} q^{i\alpha}={\cal
Q}^{i\alpha}|, \qquad \eta^i= -iD^i \left( \frac{2}{{\cal
Q}^{j\alpha}{\cal Q}_{j\alpha}}\right)|,\; \bar\eta{}_i= -i\bD_i
\left( \frac{2}{{\cal Q}^{j\alpha}{\cal Q}_{j\alpha}}\right)|, \ee
and it does not contain any auxiliary components
\cite{hyper1,hyper2,hyper3,hyper4,IL}.

One may easily check that if we define the composite superfield
$X$ as \be\label{QX} X=\frac{2}{Q^{i\alpha}Q_{i\alpha}}, \ee then
it will obey \p{X} in virtue of \p{Q} \cite{ikl1}. For the
hypermultiplet ${\cal Q}^{i\alpha}$ we defined the fermionic
components to coincide with those present in the $X$ superfield
\p{compX}, while the former auxiliary components $A_{ij}$ are now
expressed via the components of ${\cal Q}^{i\alpha}$ as
\be\label{AQ} A_{ij}=-\frac{4\im}{(q^{k\beta} q_{k\beta})^2}\left(
{\dot q}{}^\alpha_i q_{j\alpha}+{\dot q}{}^\alpha_j
q_{i\alpha}\right)- \frac{(q^{k\beta} q_{k\beta})}{2} \left(
\eta_i \bar\eta_j+\eta_j\bar\eta_i\right). \ee As in the case of
the tensor supermultiplet, one may write the full action with the
hypermultiplet self-interacting part $S_q$ added as
\be\label{Qpsi} S=S_q+S_c=-\frac{1}{32}\int dt d^4\theta {\cal
F}({\cal Q}) +S_c, \ee where now ${\cal F}({\cal Q})$ is an
arbitrary function of ${\cal Q}{}^{i\alpha}$. The action \p{Qpsi}
describes the motion of an isospin particle on a conformally flat
four-manifold carrying the non-Abelian field of a BPST instanton
\cite{sutulin2}. This system has been recently obtained in
different frameworks in \cite{Kon1,Kon2}.

To close this Section one should mention that, while dealing with
the tensor supermultiplet ${\cal V}^{ij}$ and the hypermultiplet
${\cal Q}^{i\alpha}$, the structure of the action $S_c$
\p{actionX} can be generalized to be \cite{sutulin2}
\be\label{actionY} S_c=-\frac{1}{32}\int dt
d^4\theta \; Y \Psi^\halpha {\bar\Psi}_\halpha,
\ee
with $Y$ obeying \bea\label{genY}
\triangle_3 Y = 0 & & \mbox{   in case of the tensor supermultiplet} \nn \\
\triangle_4 Y = 0 & & \mbox{   in case of the hypermultiplet}.
\eea Here $\triangle_n$ denotes the Laplace operator in a flat
Euclidean $n$-dimensional space. Clearly, our choice $Y=X+g$  with
$X$ defined in \p{XV}, \p{QX} corresponds to spherically-symmetric
solutions of \p{genY}. \setcounter{equation}0
\section{From the hypermultiplet to the tensor supermultiplet and back}
One of the most attractive features of our approach is the unified
structure of the action $S_c$ \p{actionX} which has the same form
for any type of supermultiplets, which we are using to construct a
composite superfield $X$. Just this opens the way to relate the
different systems via duality transformations. Indeed, it has been
known for a long time \cite{GR, Tolik, FG, FG1, root, DI1} that in
one dimension one may switch between supermultiplets with a
different number of physical bosons, by expressing the auxiliary
components through the time derivative of physical bosons, and
vice versa. Here we will use this mechanism to obtain the action
of the tensor multiplet\p{Vpsi} from the hypermultiplet \p{Qpsi}
one and then, alternatively, the action \p{Qpsi} (with some
restrictions) from \p{Vpsi}. In what follows, to make some
expressions more transparent, we will use, sometimes,  the
following stereographic coordinates for the bosonic components of
hypermultiplet \p{compQ}
 and tensor supermultiplet \p{v}: \bea
 && q^{11}=\frac{e^{\frac{1}{2}(u-\im \phi)}}{\sqrt{1+\Lambda\bLam}}\Lambda,\quad q^{21}=-\frac{e^{\frac{1}{2}(u-\im \phi)}}{\sqrt{1+\Lambda\bLam}},\quad q^{22} =\left( q^{11}\right)^\dagger, \quad q^{21} =-\left( q^{12}\right)^\dagger,
 \label{stq} \\
 && V^{11}=2 \im \frac{e^u}{1+\Lambda\bLam} \Lambda,\qquad V^{22} = -2 \im \frac{e^u}{1+\Lambda\bLam} \bLam,\qquad
 V^{12}=-\im e^u \left( \frac{1-\Lambda\bLam}{1+\Lambda\bLam}\right). \label{stv}
\eea
One may easily check  that these definitions are compatible with \p{XV} and \p{QX}. \vspace{0.5cm}

\noindent{\bf From hypermultiplet to tensor supermultiplet}\\
The main ingredient to get the tensor supermultiplet action from
the hypermultiplet one is provided by the expression for
``auxiliary'' components $A^{ij}$ in terms of $V^{ij}$ \p{AV} and
$Q^{i\alpha}$ \p{AQ}. Identifying the right hand sides of \p{AV}
and \p{AQ} one may find the expression of the auxiliary component
$A$ present in the superfield $V^{ij}$ in terms of components of
$Q^{i\alpha}$: \be\label{AfromQ} A=\im \left( {\dot q}{}^{i1}
q_i^2 +  {\dot q}{}^{i2} q_i^1 \right) +\frac{1}{4}
(q^{k\alpha}q_{k\alpha})^2\; q^{i1}q^{j2} \left(
\eta_i\bar\eta_j+\eta_j\bar\eta_i\right). \ee Another way, and
probably the easiest one, to check the validity of \p{AfromQ} is
to use the following superfield representation for the tensor
supermultiplet \cite{IL}: \be\label{VfromQ} V^{ij}=\im \left(
Q^{i1}Q^{j2}+Q^{j1}Q^{i2}\right). \ee This ``composite''
superfield $V^{ij}$ automatically obeys \p{V} as a consequence of
\p{Q}.

Being partially rewritten in terms of components \p{stq}, the
expression \p{AfromQ} reads \be\label{phidot} \dot\phi
=e^{-u}A-\im \frac{\dot\Lambda \bLam
-\Lambda{\dot\bLam}}{1+\Lambda\bLam}- \frac{1}{4}e^{-u}
(q^{k\alpha}q_{k\alpha})^2\; q^{i1}q^{j2} \left(
\eta_i\bar\eta_j+\eta_j\bar\eta_i\right). \ee Thus, we see that,
in order to get the action for the tensor supermultiplet, one has
to replace, in the component action for the hypermultiplet, the
time derivative of the field $\phi$ by the combination in the
r.h.s. of \p{phidot}, which includes the new auxiliary field $A$.
An additional restriction comes from the $S_q$ part of the action
\p{Qpsi}, which has to depend now only on the ``composite''
superfield $V^{ij}$ \p{VfromQ}. If it is so, then in the full
action \p{Qpsi} the component $\phi$ will enter only through
$\dot\phi$, and the discussed replacement will be
valid.\vspace{0.5cm}

\noindent{\bf From the tensor supermultiplet to the hypermultiplet}\\
It is clear that the backward procedure also exists. Indeed, from
\p{rel1} and \p{AQ} one may get the following expression for $A$:
\be\label{AfromV} A=\frac{1}{f}\left[ \dot\phi +\frac{\partial
}{\partial v_a} f (\lambda \sigma^a \bar\lambda) -B_a{\dot
v}{}_a\right], \ee where \be\label{fH} f=\frac{1}{|v|},\quad
\mbox{ and },\quad B_1=-\frac{v_2(v_3+|v|)}{(v_1^2+v_2^2)|v|},\;
B_2=\frac{v_1(v_3+|v|)}{(v_1^2+v_2^2)|v|},\; B_3=0. \ee It is easy
to check, that in the coordinates \p{v}, \p{stv} we have \be
B_a{\dot v}_a = -\im \frac{\dot\Lambda \bLam
-\Lambda{\dot\bLam}}{1+\Lambda\bLam}\quad \mbox{ and } \quad
|v|=e^u \ee in full agreement with \p{phidot}. Thus, to get the
hypermultiplet action \p{Qpsi} from that for the tensor
supermultiplet \p{Vpsi}, one has  to dualize the auxiliary
component $A$ into a new physical boson $\phi$ using \p{AfromV}.
Of course, we do not expect to get the most general action for the
hypermultiplet interacting with the isospin-containing
supermultiplet $\Psi$, because the $S_v$ part in \p{Vpsi} depends
only on the ${\cal V}$ supermultiplet. But we will get for sure a
particular class of hypermultiplet actions with one isometry, with
the Killing vector $\partial_\phi$.

\setcounter{equation}0
\section{Hyper-K\"{a}hler sigma model with isospin variables}
The consideration we carried out in the previous Section has one
subtle point. Indeed, if we rewrite \p{fH} as
\be\label{fg1}
\dot\phi = B_a {\dot v}{}_a - f_{,a} (\lambda \sigma^a
\bar\lambda)+fA, \qquad f_{,a}\equiv \frac{\partial}{\partial v_a}
f,
\ee
then the r.h.s. of \p{fg1} has to transform as a full time
derivative under supersymmetry transformations \p{Vtr}. One may
check that it is so, if $f$ and $B_a$ are chosen as in \p{fH}. But
this choice in not unique. It has been proved in \cite{AS} that
the r.h.s. of \p{fg1} transforms as a full time derivative, if the
functions $f$ and $B_a$ satisfy the equations \be\label{hkcond}
\triangle_3 f \equiv f_{,aa} = 0, \qquad f_{,a}=\epsilon_{abc}
B_{c,b}. \ee Thus, one may construct a more general action for
four-dimensional \Nf supersymmetric mechanics using the component
action for the tensor supermultiplet and substituting there the
new dualized version of the auxiliary component $A$ \p{fg1}.

Integrating over theta's in \p{Vpsi} and eliminating the auxiliary
fermions $\rho^{\halpha}$ \p{ad5a}, \p{rel1}, we will get the
following component action for the tensor supermultiplet:
\bea\label{hk}
S&=&\frac{1}{8} \int dt \left[ F \left( {\dot
v}_a{\dot v}_a +A^2\right) +\im \left( {\dot\xi}{}^i
\bxi_i-\xi^i\dot{\bxi}_i\right)+\im \epsilon_{abc}
\frac{F_{,a}}{F}{\dot v}_b\Sigma_c -
\im \frac{F_{,a}}{F}\Sigma_a A -\frac{1}{6} \frac{\triangle_3 F}{F^2}\Sigma_a\Sigma_a \right. \nn \\
&& -2 \im \left( {\dot w}{}^i{\bar w}_i -w^i{\bar w}_i\right) +4 \frac{1+3 g |v|}{F (1+g |v|)^2 |v|^4}
\left(v_a I_a\right)\left(v_b\Sigma_b\right)- 4\frac{g}{F (1+g|v|)^2 |v|}\left( I_a \Sigma_a\right) \nn \\
&& \left. -\frac{4\im}{(1+g|v|)|v|^2}\left(v_a I_a\right)\; A+
\frac{4\im}{(1+g|v|)|v|^2}\epsilon_{abc}v_a{\dot v}_b I_c \right],
\eea where \be\label{defs1} F=\triangle_3\; {\cal F}({\cal V})|,
\qquad I^a=\frac{\im}{2}\left( w \sigma^a  {\bar w}\right), \qquad
\Sigma^a =-\im \left( \xi \sigma^a  {\bar \xi}\right),
\ee
and the re-scaled  fermions and isospin variables are chosen to be
\be
\xi^i = \sqrt{F}\; \lambda^i, \qquad w^i
=\sqrt{g+\frac{1}{|v|}}\;u^i.
\ee
Substituting \p{fg1} into
\p{hk}, we obtain  the resulting action
\bea\label{hkf} S&=&
\frac{1}{8} \int dt \left[ F \left( {\dot v}_a{\dot v}_a+
\frac{1}{f^2}\left( \dot\phi - B_a {\dot v}_a\right)^2 \right)+\im
\left( {\dot\xi}{}^i \bxi_i-\xi^i\dot{\bxi}_i\right)
-2 \im \left( {\dot w}{}^i{\bar w}_i -w^i{\bar w}_i\right) \right. \nn \\
&& -\im \left[ \frac{1}{f}\delta_{ab}\left(\dot\phi-B_c{\dot v}_c\right)+\epsilon_{abc}{\dot v}_c\right]
\left( \frac{F_{,a}}{F}\Sigma_b +\frac{4}{(1+g |v|)|v|^2} v_a I_b \right) \nn \\
&& +\frac{4}{F} \frac{1+3 g |v|}{(1+g|v|)^2 |v|^4} \left(v_a I_a\right) \left( v_b \Sigma_b\right)-
\frac{1}{F} \frac{4g}{(1+g |v|)^2 |v|}\left( I_a \Sigma_a\right) \nn \\
&& \left. +\frac{1}{3 F^2}\left( \frac{F_{,a}f_{,a}}{f} -\frac{F
f_{,a}f_{,a}}{f^2}- \frac{1}{2} \triangle_3 F \right) \Sigma_b
\Sigma_b \right].
\eea
The action \p{hkf} is our main result. It
describes a motion of a \Nf supersymmetric four-dimensional isospin carrying
particle in a non-Abelian field of
some monopole. The metric of this four-dimensional space is
defined in terms of two functions: the bosonic part of our
pre-potential $F$ \p{defs1} and the harmonic function $f$
\p{hkcond}. The supersymmetric version of the coupling with the
monopole (second line in the action  \p{hkf}) is defined by the
same harmonic function $f$ and the coupling constant $g$. In the
more general case \p{actionY}, we will have two harmonic functions
- $f$ and $Y$, besides the pre-potential $F$.

Among all possible systems with the action \p{hkf} there is a very
interesting sub-class which corresponds to hyper-K\"{a}hler sigma
models in the bosonic sector. This case is distinguished by the
condition \be\label{HK} F=f. \ee Clearly, in this case the bosonic
kinetic term of the action \p{hkf} acquires the familiar form of
the one dimensional version of the general Hawking-Gibbons
solution for four-dimensional hyper-K\"{a}hler metrics with one
triholomorphic isometry \cite{GH}: \be\label{GH}
S_{kin}=\frac{1}{8} \int dt \left[ f {\dot v}_a{\dot v}_a+
\frac{1}{f}\left( \dot\phi - B_a {\dot v}_a\right)^2
\right],\qquad \triangle_3 f=0,\; \mbox{rot } \vec{B}
=\vec{\nabla} f. \ee It is worth to note that the bosonic part of
\Nf supersymmetric four dimensional sigma models in one dimension
does not necessarily have to be a hyper-K\"{a}hler one. This fact
is reflected in the arbitrariness of the pre-potential $F$ in the
action \p{hkf}. Only under the choice $F=f$ the bosonic kinetic
term is reduced to the Gibbons-Hawking form \p{GH}. Let us note that
for hyper-K\"{a}hler cases the four-fermionic term in the action
\p{hkf} disappears. This fact has been previously established in
\cite{AS}. Now we can see that the additional interaction with
background non-Abelian gauge field does not destroy these nice
properties.

Among all possible bosonic metrics one may easily find the following interesting ones.\vspace{0.5cm} \\
\noindent{\bf Conformally flat spaces.}\\
There are two choices for the function $f$ which correspond to the conformally flat metrics in the bosonic sector.

The first choice is realized by
\be\label{1}
f=\frac{1}{|v|}.
\ee
This is just the case we have considered in Section 2. Tee gauge field in this case is the field
of BPST instanton \cite{sutulin2}.

Next, an almost trivial solution, also corresponding to the flat metrics in the bosonic sector, is selected by the condition
\be\label{2} f= const, \qquad B_a = 0. \ee
Note, that the relation with the
tensor supermultiplet, in this case, is achieved through the
following ``composite'' construction of $V^{ij}$ \cite{ILS}
\be\label{2v} V^{ij}= {\cal Q}^{(i\alpha)}. \ee
One may check that
the constraints on $V^{ij}$ \p{V} directly follow from \p{2v} and
\p{Q}.

Let us remind that in both these cases we have not specified the
pre-potential $F$ yet. Therefore, the full metrics in the bosonic
sector is defined up to this function.
\vspace{0.5cm} \\
\noindent{\bf Taub-NUT space.} \\
One should stress that the previous two cases are unique, because
only for these choices of $f$  the resulting action \p{hkf} can be
obtained directly from the hypermultiplet action \p{Qpsi}. With
other solutions for $f$ we come to the theory with the nonlinear
\Nf hypermultiplet \cite{AS,nelin1}. Among the possible solutions
for $f$ which belongs to this new situation the simplest one
corresponds to one center Taub-NUT metrics with \be\label{3}
f=p_1+\frac{p_2}{|v|}, \qquad p_1,p_2=\mbox{const}. \ee In order
to achieve the maximally symmetric case, we will chose these
constants as
\be\label{3a} p_1=g, \quad p_2 =1 \quad \rightarrow
\quad f=g+\frac{1}{|v|}.
\ee
With such a definition $f$ coincides
with the function $Y=g+\frac{1}{|v|}$ \p{actionY} entering in our
basic action $S_c$ in \p{actionX}, \p{XV}. To get the Taub-NUT
metrics, one has also to fix the pre-potential $F$  to be equal to
$f$. The resulting action which describes the \Nf supersymmetric
isospin carrying particle moving in a Taub-NUT space reads
\bea\label{TN}
S_{Taub-NUT}&=& \frac{1}{8} \int dt \left[  \left(
g+\frac{1}{|v|}\right)  {\dot v}_a{\dot v}_a+ \frac{1}{\left(
g+\frac{1}{|v|}\right)}\left( \dot\phi - B_a {\dot v}_a\right)^2
+\im \left( {\dot\xi}{}^i \bxi_i-\xi^i\dot{\bxi}_i\right)
-2 \im \left( {\dot w}{}^i{\bar w}_i -w^i{\bar w}_i\right) \right. \nn \\
&& +\frac{\im}{(1+g |v|)|v|^2} \left[ \frac{v_a}{\left( g+\frac{1}{|v|}\right)}\left(\dot\phi-B_c{\dot v}_c\right)-\epsilon_{abc}v_b{\dot v}_c\right]
\left( \Sigma_a -4 I_a \right) \nn \\
&&\left. +\frac{4(1+3 g |v|)}{(1+g|v|)^3 |v|^3} \left(v_a I_a\right) \left( v_b \Sigma_b\right)-
 \frac{4g}{(1+g |v|)^3}\left( I_a \Sigma_a\right) \right].
\eea The bosonic term in the second line of this action can be
rewritten as \be\label{inst} {\cal A}_a I_a =\frac{\im}{2}\left[
\frac{1}{f}\; \frac{\partial \log f}{\partial v_a}\left( \dot\phi
- B_a {\dot v}_a\right)-\epsilon_{abc} \frac{\partial \log
f}{\partial v_b}{\dot v}_c\right] I_a ,
\ee
where $f$ is defined
in \p{3a}. In this form the vector potential ${\cal A}_a$
coincides with the potential of a Yang-Mills $SU(2)$ instanton in
the Taub-NUT space \cite{inst1,inst2}, if we may view $I_a$, as
defined in \p{defs1}, as proper isospin matrices. The remaining
terms in the second and third lines of \p{TN} provide a \Nf
supersymmetric extension of the instanton.

Finally, to close this Section, let us note that more general
non-Abelian backgrounds can be obtained from the multi-centered
solutions of the equation for the harmonic function $Y$ \p{genY},
which defined the coupling of the tensor supermultiplet with
auxiliary fermionic ones.
Thus, the variety models we
constructed are defined through three functions: pre-potential ${\cal F}$ \p{Vpsi} which is an arbitrary function,
3D harmonic function $Y$ \p{actionY}, \p{genY} defining the coupling with isospin variables and, through,
again 3D harmonic, function $f$ \p{fg1}, \p{hkcond} which appeared during the dualization of the auxiliary
component of the tensor supermultiplet.
 It is clear that
we can always redefine $F$ to be $F={\tilde F}f$. Thus, all our
models are conformal to hyper-K\"{a}hler sigma models with \Nf
supersymmetry describing the motion of a particle in the
background non-Abelian field of the corresponding instantons.

\setcounter{equation}0
\section{Conclusion}

In the present paper we constructed the Lagrangian formulation of
\Nf supersymmetric mechanics with hyper-K\"{a}hler sigma models in
the bosonic sector in the non-Abelian background gauge field. The
resulting action includes the wide class of \Nf supersymmetric
mechanics describing the motion of an isospin-carrying particle
over spaces with non-trivial geometry. In two examples we
discussed in details, the background fields are identified with
the field of BPST instantons in the flat and Taub-NUT spaces.

The approach we used in the paper utilized two ideas: (i) the
coupling of matter supermultiplets with an auxiliary fermionic
supermultiplet $\Psi^{\halpha}$ containing on-shell four physical
fermions and four auxiliary bosons playing the role of isospin
variables and (ii) the dualization of the auxiliary component $A$
of the tensor supermultiplet into a fourth physical boson. The
final action we constructed contains three arbitrary functions:
the pre-potential ${\cal F}$, a 3D harmonic function $Y$ which
defines the coupling with isospin variables and, again 3D
harmonic, a function $f$ which appeared during the dualization of
the auxiliary component of the tensor supermultiplet. The
usefulness of the proposed approach is demonstrated by the
explicit example of the simplest system with non-trivial geometry
- the \Nf supersymmetric action for one-center Taub-NUT metrics.
We identified the background gauge field in this case, which
appears automatically in our framework, with the field of the BPST
instanton in the Taub-NUT space. Thus, one may hope that the other
actions will possess the same structure.

Of course, the presented results are just preliminary in the full
understanding of \Nf supersymmetric hyper-K\"{a}hler sigma models
in non-Abelian backgrounds. Among interesting, still unanswered
questions, one should note the following ones:
\begin{itemize}
\item The full analysis of the general coupling with an arbitrary harmonic function $Y$ has yet to be
carried out.
\item The structure of the background gauge field has to be further clarified: is this really the field of some
monopole (instanton) for some hyper-K\"{a}hler metrics?
\item The Hamiltonian construction is really needed. Let us note that the Supercharges have to be very
specific, because the four-fermions coupling is absent in the case
of HK metrics!
\item  It is quite interesting to check the existence of  the conserved Runge-Lenz vector in the
fully supersymmetric version.
\item The explicit examples of other hyper-K\"{a}hler metrics (say, multi-centered Eguchi-Hanson and Taub-NUT ones) would be
very useful.
\item The questions of quantization and analysis of the spectra, at least in the cases of well known, simplest
hyper-K\"{a}hler metrics, are doubtless urgent tasks.
\end{itemize}

Finally, let us stress that our construction is restricted to the
case of hyper-K\"{a}hler metrics with one translational
(triholomorphic) isometry. It will be very nice to find a similar
construction applicable to the case of geometries with rotational
isometry. We hope this may be done within the approach discussed
in \cite{tri}.

 \setcounter{equation}0
\section*{Acknowledgements}
We thank Andrey Shcherbakov for useful discussions. This work was
partially supported by the grants RFBF-09-02-01209 and
09-02-91349, by Volkswagen Foundation grant~I/84 496 as well as by
the ERC Advanced
Grant no. 226455, \textit{``Supersymmetry, Quantum Gravity and Gauge Fields''%
} (\textit{SUPERFIELDS}).

\end{document}